\documentclass[preprint,12pt]{elsarticle}
\usepackage[T1]{fontenc}
\usepackage{siunitx}
\usepackage{algorithm}
\usepackage[noend]{algpseudocode}
\usepackage{array}
\usepackage[caption=false,font=normalsize,labelfont=sf,textfont=sf]{subfig}
\usepackage{textcomp}
\usepackage{stfloats}
\usepackage{url}
\usepackage{verbatim}
\usepackage{graphicx}
\usepackage{grffile}
\usepackage{balance}

\algnewcommand{\TRUE}{\textbf{True}}
\algnewcommand{\FALSE}{\textbf{False}}

\DeclareSIUnit{\belmilliwatt}{Bm}
\DeclareSIUnit{\dBm}{\deci\belmilliwatt}
\usepackage{amssymb}
\journal{Computer Networks}

\begin{document}

\begin{frontmatter}



\title{Routing and wavelength assignment in quantum key distribution networks: power control heuristics for quantum-classical multiplexing.}


\author[inst1,inst2]{Lidia Ruiz}

\affiliation[inst1]{organization={Departamento de Teoría de la Señal y Comunicaciones e Ingeniería Telemática, Universidad de Valladolid},%Department and Organization
            addressline={Paseo Belén 15}, 
            city={Valladolid},
            postcode={47011}, 
            country={Spain}}

\affiliation[inst2]{organization={Laboratory for Disruptive Interdisciplinary Science (LaDIS), Universidad de Valladolid},%Department and Organization
            city={Valladolid},
            postcode={47011}, 
            country={Spain}}

\author[inst1,inst2]{Juan Carlos Garcia-Escartin}

\begin{abstract}
	As quantum key distribution networks grow in size and complexity, resource assignment has become increasingly important. In passive optical networks without wavelength conversion, we need to assign a full route between origin and destination with the same wavelength from a finite set. This problem is computationally intensive and the common solution in classical optical networks is using heuristics. We adapt these heuristics to hybrid quantum networks where the quantum channel can share some of the optical links with classical channels. In this quantum-classical multiplexing, nonlinear effects can become the limiting factor in the range of the network. The signal in the classical channels can be subject to Raman Scattering or Four-Wave-Mixing and produce light in the quantum channels. While these processes are not efficient, even a single photon can ruin the quantum channel. We propose heuristics for the routing and wavelength assignment problem for hybrid quantum-classical networks with power control for the classical channels. By keeping the transmitted power to its bare functional minimum, we can reduce the interference to the quantum channels. We study their efficiency under different scenarios.
\end{abstract}




\end{frontmatter}


\section{Quantum key distribution networks}
	Communication protocols that use the laws of Quantum Mechanics allow new applications beyond classical possibilities. In particular, the limitations of quantum measurement, where the state of a superposition is destroyed after being observed, can be used to establish an arbitrarily long secret key between two parties that use a private, but insecure, quantum communication channel and a public, but authenticated, classical communication channel. This offers an additional level of security based on physical principles that can be combined with the usual cryptographic methods based on computational complexity. Quantum Key Distribution, QKD, is a mature quantum technology for which there are multiple protocols, experimental proofs of concept, standards and commercial products \cite{KLH15,MGM15,BAL17,XMZ20}. Now that QKD links between two sides can be deployed with relative ease, there is a growing interest in QKD networks with multiple nodes in which any pair of nodes can establish a secret key securely. 
	
	As in most quantum communication demonstrations, in order to cover large distances fast, photons are the quantum system of choice. The main proposals are free-space QKD, including satellite communication, and optical fiber QKD networks that take advantage of the existing optical communication network infrastructure. We will concentrate on these fiber networks.
	
	While the first QKD demonstrations used dedicated dark fiber links, the current objective is a gradual integration with existing fiber communication networks. Ideally, QKD networks would be part of flexible Software Defined Networks, SDN, where there is a logical control plane deciding which connections can be established and how the network resources are assigned \cite{AHH17,CZW19,MAB19}. 
	
	QKD links need a classical channel for public discussion, including error correction and privacy amplification, and a quantum channel which can basically use most of the existing classical optical fiber technology with the important exception of amplifiers, which must be considered with care. The same principles that prevent an eavesdropper learning a quantum state forbid noiseless amplification and there is a distance limit for QKD links. Photons can get lost during transmission and, above a certain loss threshold, we can no longer guarantee the observed imperfections are not due to an adversary that is attacking the system. Current QKD systems, depending on the quality of the detectors and the optical fiber used, can cover distances from tens to hundreds of kilometers. The systems that use standard optical fiber devices, with simple transmitters and detectors, are limited to the smaller end of that range. 
	
	QKD hardware is chosen to minimize noise and losses. We will concentrate on discrete variable QKD protocols where the signal sent into the channel usually has less than one photon on average. This makes eavesdropping harder as we cannot measure the photon without disturbing its state and we cannot keep a copy for measurement and forward another photon. However, this means that stray photons in the quantum channel band can introduce fatal errors. The heuristics we give could also be considered for continuous variable QKD protocols \cite{ZBL24} where amplifiers can help in certain situations. We will discuss only discrete variable protocols, which need a higher protection against photon contamination.
	
	Thermal photons are not much of an issue compared to the noise photons that appear in hybrid networks where the same optical fiber carries the classical and the quantum channels needed in QKD or when we share the resources with classical traffic. In these cases, there can be contamination from the classical signal. For wavelength division multiplexing, the main contamination mechanism is the presence of nonlinear effects that convert photons from the classical channel bands into the quantum band. In particular, depending on the chosen wavelengths, Raman Scattering and Four-Wave-Mixing can create noise photons from classical channels with moderate powers with a high enough probability to be the limiting factor to the maximum link length \cite{CTP09,PTC09,EWL10,FXT14,GSB21}. These processes are more important for longer shared paths between the classical and the quantum signals.
	
	We propose routing heuristics that assign paths and wavelengths to the classical and quantum channels taking this contamination into account. The algorithms are inspired by the power control methods used in wireless sensor networks to reduce the consumed power and make battery lives longer \cite{GOW07,HMJ09}.

	\section{Network description and traffic demands. Routing and wavelength assigment}
	The most natural way to represent optical networks is using graphs where each end user is at the nodes and optical fiber links are represented as the links, or edges, of the graph. We assume these links can carry a maximum total number of wavelengths $W_T$. 
	
	We consider a series of network demands where we are given an origin and a destination which ask for a \emph{lightpath}: a path connecting the origin and destination nodes using a dedicated wavelength. Two lightpaths sharing a link must have a different wavelength, but non-overlapping lightpaths can use the same or different wavelengths. These restrictions model what happens in optical networks without wavelength conversion, which is a reasonable scenario for our QKD network. Wavelength conversion below the single photon level is technologically challenging and, while it can be considered for the classical signals, it would increase the complexity and cost of the QKD network.
	
	We are interested in optimal resource assignment so that the limited number of transmitters and receivers at each node and the fibers in our network can fulfill the maximum possible number of traffic requests.
	
	For a given topology and a series of traffic requests, the routing and wavelength assigment, RWA, problem consists in finding lighpaths for each origin and destination, minimizing the number of requests that cannot be fulfilled (for instance if there are no available wavelenghts in any route between the nodes). This problem is known to be NP-complete for its usual formulation \cite{CGK92}. Full optimization is quite involved and, in practice, there is a set of efficient heuristics which approximate the best solution faster and using less computational resources \cite{ZJM00}.
	
	A widely used strategy is using the heuristic shortest-path-first-fit, SPFF, and its variations. In this heuristic, we begin using Dijkstra's algorithm to find the shortest path between origin and destination (or the $k$ shortest paths in many cases). Then, assuming we have an ordered list of available wavelengths, we assign a lightpath to the shortest route which has a free wavelength, assigning the wavelength with the smallest index in our ordering. If no such a path can be found, the connection is \emph{blocked} (the traffic request is denied).
	
	The RWA and related problems are also relevant to QKD networks and there is an increasing interest in resource allocation methods for quantum networks \cite{CZW22}, including schemes with trusted or semitrusted relay nodes \cite{DKZ20,SPB21,YLZ22,CZZ22,CZW22,BMD23,CCCZ23,ZAG23} and with combined time and wavelength multiplexing \cite{CZY17,NZY17,CYZ18,ZCW18,DKZ20,SPB21,YLZ22,ZAG23}. We continue here the line started in \cite{RG23}, which also includes a comparison to other RWA proposals for QKD. The proposed heuristics focus on reducing the problems caused by multiplexing classical and quantum channels. The ultimate purpose is to give methods \emph{as simple as possible which can be deployed with little changes in current optical fiber networks}. The main difference with respect to recent quantum-specific RWA algorithms is that our heuristics explicitly take into account the most limiting physical noise mechanisms in QKD and provide mitigation at the network level. This naturally leads to power control strategies for the classical channels.	
	
	\section{Link model: nonlinear effects, signal-to-noise ratios and quantum key rate}
\label{LinkModel}
	When routing classical and quantum channels through the same optical fiber, we need to estimate how much noise the classical channels create in the quantum channels. We will use a simplified model for each link that gives an approximate measure of the expected interference instead of a complete, and costly, physical simulation including all the fiber parameters and with accurate models of all the relevant nonlinear effects. This way, the heuristics can approximate the limiting noise effects in quantum-classical multiplexing while they are still simple and use few computational resources.
	
	The quantum channels in a QKD network have new specific restrictions. Even if we have all the necessary resources, any quantum lightpath with excessive losses or noise will not be useful for key distribution. Above a certain Quantum Bit Error Rate, QBER, QKD protocols can no longer guarantee the generated keys are private. This limit is usually given as a function of distance, which determines the channel loss. The concrete function relating losses and the maximum achievable key rates depends on the particular protocol used, but there is always a distance above which the quantum communication phase is useless. 
	
	We will use our basic noise estimations to determine a new blocking criterion for the quantum channels. A quantum request can be blocked because there are no available wavelengths, just as classical channels are, but we will consider that quantum channels that cannot generate secret key bits above a certain rate must be blocked to avoid them hogging the network resources without a useful outcome.
	
	The effect of the noise will be estimated with the quantum signal-to-noise ratio, QSNR, in the quantum channel. In most natural scenarios, a smaller signal-to-noise ratio in the quantum channel is related to a higher QBER and a lower established key rate. We will use a QSNR threshold to determine whether a quantum channel can provide fresh secret bits to a key or not. For each QKD protocol, we can always find out which QSNR threshold corresponds to an exchange that is doomed to fail due to a QBER above the minimum value. In general, this is a direct function of the total distance of the quantum link, but, in hybrid QKD networks, the noise produced by the classical part is also relevant. We will show the effects on the final key rate with an example for the BB84 protocol under certain restrictions.
	
	\subsection{Nonlinear effects}
	A common solution for quantum-classical multiplexing is keeping the classical channels in the C-band around \SI{1550}{\nano\meter} and moving the quantum transmission to the O-band around \SI{1310}{\nano\meter}. We will work with this channel arrangement, for which the dominant noise source will be Raman scattering \cite{PTC09}. We will be concerned by the tails of the Raman spectrum produced by signals in the C-band. These tails have weak levels in the O-band, but they are noticeable at the quantum level. 

Raman scattering occurs when the photons of the classical channel interact with the vibrational modes of the optical fiber. There can be both spontaneous and stimulated Raman scattering processes, but, for our channel distribution and power levels, we just need to worry about spontaneous Raman scattering. Spontaneous Raman scattering, in turn, can present Stokes and anti-Stokes processes. In Stokes processes a photon produces a lower energy photon and a phonon. In anti-Stokes processes, a photon and an existing phonon in the fiber produce a new photon with a higher energy (a lower wavelength). The pump laser signal and the scattered photon can go in the same direction (copropagation) or in opposite directions (counterpropagation). We are considering unidirectional fiber links, where duplex communication is done with one fiber for each direction, and with a pump, the classical channel, at a much higher wavelength than the quantum channels. In that scenario, the most important photon contamination in the quantum channel are the photons coming from anti-Stokes spontaneous Raman scattering of the photons of the classical channel copropagating with the quantum channel. For our frequencies and channel powers, the Raman noise power can be approximated as \cite{CTP09,PTC09}:
\begin{equation}
\label{RamanFormula}
\text{RNP}=P_0 \frac{\beta_q}{\alpha_q-\alpha_c}\left( e^{-\alpha_c z}-e^{-\alpha_q z}\right),
\end{equation}
where $P_0$ is the power of the classical signal at the input of the fiber, $\alpha_c\neq  \alpha_q$ are the attenuation constants, in \unit{\kilo\meter^{-1}}, for our optical fiber at the pump (classical) and signal (quantum) wavelengths and $\beta_q$ is the spontaneous Raman scattering coefficient for the signal (pump to signal scattering) in \unit{\kilo\meter^{-1}}. In general, we have a function $\beta_q=\beta(\lambda_s,\lambda_p,\delta \lambda)$ which depends on the wavelengths of the signal and the pump and the bandwidth we are observing. Raman scattering is less efficient for large frequency shifts between pump and signal (high $\lambda_s-\lambda_p$) and, in the tail at the O-band where we have our quantum channnels, we can consider a constant function for the bandwidth of interest. We consider a receiver with a filter of bandwidth $\delta \lambda$ so that $\beta_q=\beta\delta\lambda$. 

In the standard silica fiber we consider, we have an attenuation $\alpha_{cdB}=\SI{0.17}{\deci\bel/\kilo\meter}$, corresponding to $\alpha_c=\SI{0.0391}{\kilo\meter^{-1}}$ for the classical channels around \SI{1550}{\nano\meter} and $\alpha_{qdB}=\SI{0.32}{\deci\bel/\kilo\meter}$, $\alpha_q=\SI{0.0737}{\kilo\meter^{-1}}$, for the quantum channels around \SI{1310}{\nano\meter}.

In Equation (\ref{RamanFormula}), we can rewrite the term giving the evolution with distance $e^{-\alpha_c z}-e^{-\alpha_q z}$ as $e^{-\alpha_c z}\left(1-e^{-(\alpha_q-\alpha_c) z}\right)$. For our attenuation constants and total distances above \SI{50}{\kilo\meter}, the second exponential introduces a correction in the Raman power below the 20\%. Depending on the scenario, we could, for simplicity, drop the second exponential in the model to give an upper bound for the Raman noise, but the term can be included if there are many short links. This approximation could be useful if we wanted to collect multiple nonlinear effects in one term. We could then plug all the terms into one constant and reduce the computation complexity when approximated linearizations give a good enought estimation. As the computational cost is small for the Raman formula we have discussed, we will keep the full term. 

For multiple channels we can just sum the contribution of each classical pump field \cite{FXT14} assuming that we have a constant attenuation and spontaneous Raman scattering coefficient for all the signals in the C band. In the final term, nonlinear contamination for each different optical fiber can be reduced to a constant $\gamma_{nl}$ multiplied by the input power and the attenuation exponentials. If needed, the model can also be easily adapted to other scenarios by adding a term for the counterpropagating signals or distinct values of $\gamma_{nl}$ for heterogeneous networks with fibers with different nonlinear responses.

	\subsection{Classical and quantum signal to noise ratios}
	We will use the signal-to-noise ratio of the classical and quantum channels as our guiding quality metric.
	
	In the classical channels we use the usual definition for the signal-to-noise ratio $SNR=P/N$ as the ratio between the signal power $P$ and the noise power $N$. For values below a certain threshold, which depends on the receivers and modulation schemes, we assume the signal cannot be interpreted correctly. If we have a fixed power at each transmitter, the receivers at the end of shorter links will receive a higher power. We are going to assume there is a fixed noise power for each link, mostly related to thermal noise at the receivers. For each channel we have 	
	\begin{equation}
	\label{SNR}
		\text{SNR} = \frac{P}{N} = \frac{e^{-\alpha_{c} L_c} P_{tx,c}}{N},
	\end{equation}
where $\alpha_{c}$ is the attenuation at the transmission wavelength in natural units (\SI{}{\kilo\meter^{-1}}), $L_c$ is the total length of the path we use for the channel, $P_{tx,c}$ is the launch power of the classical signal and $N$ is the noise term. All the signal and noise powers are normalized to a reference level of \SI{1}{\milli\watt} in the simulations.
	
	For the usual lengths in QKD, most classical channels can be transmitted at reasonable power levels. This means that, for short paths, we are actually using more power than required for correct classical communication. Apart from increasing the energy bill, this excess transmission power harms the quantum channels. 
	
	The heuristics we propose for routing will adjust the classical power at the transmitter so that there is a fixed power for any link at the receiver after a route has been chosen. The power can be adjusted to whatever SNR threshold is considered acceptable. This reduces the total transmitted power which will, in turn, reduce the number of noise photons in the classical channel.
	
	We estimate the effects of the noise on the quantum channel by defining a \emph{quantum signal-to-noise ratio} that compares the power of the quantum channel at the receiver and all the noise during detection, including the photons coming from nonlinear effects. We define the formula
	\begin{equation}
		\label{QSNR}
		\text{QSNR} = \frac{e^{-\alpha_{q} L_q} P_{tx,q}}{N_{F}+\gamma_{nl}\sum_{i} P_{tx,i}\cdot \left( e^{-\alpha_c L_{i}}-e^{-\alpha_q L_{i}}\right)}.
	\end{equation}
	Here, $\alpha_{q}$ is the attenuation of the fiber at the transmission wavelength of the quantum signal (in \SI{}{\kilo\meter^{-1}}), $L_q$ is the length of the route allocated to quantum channel $q$, $P_{tx, q}$ is the transmitted power of that quantum channel, $N_{F}$ is a fixed noise term, $\gamma_{nl}$ is a fixed nonlinear coefficient term that can be adapted to the physical parameters of the fiber used in our particular network, $P_{tx, i}$ represents the classical signal launch power at the beginning of each shared span $i$ with shared length $L_i$ and $\alpha_{c}$ represents the attenuation coefficient at the transmission wavelength of the classical channel in \SI{}{\neper/\kilo\meter}. The shared span is given by $L_i$, which corresponds to the length of the shared fiber. The sum includes all the fiber links in the path of the quantum signal in which there are classical channels. We will add one term to the sum for each classical channel sharing that link.
	
	We consider all the powers in the QSNR are normalized to the single photon level for the corresponding wavelength in the O-band so that we can interpret them as a photon number. The numerator in Equation (\ref{QSNR}) gives an estimation of the expected photon number that reaches the receiver in a detection window. It is computed from the transmitted photon number and the attenuation in the journey to the destination node. We will also include in that term effects that are equivalent to losses, like detection efficiencies below 1, which are common for the simplest single photon detectors. 
	
  Single photon detectors are a key element in discrete variable QKD. The two most important parameters are their quantum efficiency $\eta$ and their dark count rate, DCR. The detector efficiency gives the fraction of photons that reach the photon detector that are actually converted into a macroscopical electrical pulse that can be detected. Dark counts give the rate at which the detector produces a false count (a phonon or any thermal process triggers the same response as a transmitted photon would when there are none). At the usual telecom wavelengths, most affordable single photon detectors have a low detection efficiency, especially if we want to keep dark counts at bay \cite{Had09}.

Many systems use avalanche photodiodes which are notoriously inefficient at the IR telecom bands, where 10-20\% efficiencies are the norm. $P_{tx,q}$ will be set to $\mu \eta$, where $\eta$ is the quantum efficiency of the detector and $\mu$ is the average photon number at the output of a transmitter generating weak coherent states with an amplitude giving a mean photon number $\mu<1$. The value of $\mu$ can be adapted to achieve the optimal key rate for a particular protocol or can be given a fixed value, typically much smaller than one to avoid photon number splitting attacks. 
	
	The normalized noise power can be interpreted as the average number of spurious photons. We include two terms for general noise, $N_F$, and noise caused by the classical channels $N_c$.
	
	The fixed amount $N_F$ that collects thermal photons and equivalent effects, like dark counts in the detectors (false detections caused by phonons or other effects that are not related to the incoming light). In our experiments we consider the dominant fixed source of noise are dark counts, which happen in each detection bin with a probability $p_d$.
	
	We estimate $N_c$ with the simplified model we have described in Section \ref{LinkModel} and write the total noise as the product of a non-linear coefficient $\gamma_{nl}$ by the sum of the classical power of each classical channel multiplied by its corresponding exponential distance terms for the shared total distance with the quantum channel of interest.

In the general model, $\gamma_{nl}$ is left as an adjustable parameter and, for any physical QKD network, its value will depend on the detectors we use, the physical properties of the optical fiber in our network and the parameters of the QKD protocol such as time widths for detection or the transmitted number of photons.

We will give an example system as a reference to show how to set the parameters for a concrete scenario and study the changes in the achievable key rate for a specific QKD protocol, BB84. The parameters can also be studied independently if the focus is on the effect of different network topologies or more general scenarios.

In order to see how many photons are scattered into the O-band and detected at the QKD receiver we can go to Equation (\ref{RamanFormula}). The number of photons that produce a false count are
\begin{equation}
\eta \text{RNP} \frac{\Delta T}{h\nu}\approx  \eta\frac{\beta_q}{\alpha_q-\alpha_c} \left(e^{-\alpha_c z}-e^{-\alpha_q z}  \right)\frac{\Delta T}{h\nu} P_0=\gamma_{nl} P_0  \left(e^{-\alpha_c z}-e^{-\alpha_q z}  \right)
\end{equation}
where $\Delta T$ is the time of the detection window at the receiver and $\eta$ is the quantum efficiency of the detector and we divide the input noise power by the energy of a single photon at \SI{1550}{\nano\meter} to get the noise photon rate reaching the detector. 

We take a reference QKD system with a detector that, for a quantum efficiency $\eta=0.1$, has a dark count probability $p_d=10^{-4}$ per detection bin, corresponding to a time window $\Delta T =\SI{10}{\nano\second}$ and a detector with a dark count rate of \SI{10}{\kilo\hertz}. We normalize input powers so that $P_0=1$ for a \SI{1}{\milli\watt} classical signal at \SI{1550}{\nano\meter} and absorb the $7.8\cdot 10^7$ photons per time bin of such a signal into the constant $\gamma_{nl}$. We consider an optical fiber with a constant Raman coefficient for the portion of the O band we measure.

 For all these parameters and an optical fiber with a Raman coefficient $\beta_q=\beta\delta\lambda=\SI{e-12}{\kilo\meter^{-1}}$ at \SI{1310}{\nano\meter}, consistent with measured values for typical fibers with a filter of \SI{100}{\giga\hertz} \cite{LZS17}, we get $\gamma_{nl}=2.25\cdot 10^{-4}$.

This approximation gives a reasonable estimation of which classical channels are likely to introduce the most noise without a detailed physical model of each situation. The parameters can be adjusted to model each particular scenario, including using different non-linear coefficients for links with different optical fibers or when the classical and quantum signals are copropagating or counterpropagating.

	\subsection{Quantum key rate example}
	QKD exchanges start with an initial phase where the quantum states are sent through the optical fiber and measured. The results of the measurements are used to estimate the QBER for the current state of the fiber link and the devices at the receiver. After this initial estimation, there is a phase of error correction and privacy amplification where the bits from the raw key, which are directly related to the measurement results, are modified to produce a smaller secure key which should be the same for both parties. The size of this final key depends on the estimated information a potential eavesdropper can have, assuming any imperfection during the quantum state transmission and measurement has been caused by an active adversary. 
	
	The rate at which we can produce a secure key depends on the protocol, the link distance and certain parameters of the channel and the equipment. The attenuation in the optical fiber $\alpha$ and the efficiency $\eta$ and dark count rate $p_d$ of the detector are usually the most important variables.
		 
	Each QKD protocol has different security proofs that bound what an adversary can do under different assumptions. The best models that give tighter bounds and avoid throwing away more bits than necessary can be quite involved and depend on the particular QKD protocol \cite{CNN25}. There are also different numerical techniques that optimize the way the key is extracted from the measurements \cite{LPC25}. While we give a general description in terms of the QSNR, which gives a good intuition of the expected improvement for any protocol, in this Section we give a simple example of key rate values using the simplified key rate formula for the BB84 protocol under collective combined photon-number-splitting and cloning attacks as described in \cite{NSG05}. Under this scenario, we can approximate the maximum achievable secure key rate as
\begin{equation}
\label{eq:eqRBB84}
R\!=\!\frac{1}{2}\!\left(\mu t \eta + 2 p_d \right)\!\left[ 1\!-\!H(\text{QBER})\right] -\frac{1}{2}\mu t \eta\left[\! \left(t-\frac{\mu}{2}\right)\!I_1(D_1)+\frac{\mu}{2}\right]
\end{equation}
for a quantum bit error rate (QBER) 
\begin{equation}
\label{eq:QBERBB84}
\text{QBER}=\frac{1}{2}-\frac{V}{2\left(1+\frac{2p_d}{\mu t \eta}\right)},
\end{equation}
for a system using a detector with efficiency $\eta$ and a dark count probability per gate $p_d$. Here, $$I_1(D_1)=1-H\left(\frac{1}{2}+\sqrt{D_1(1-D_1)}\right)$$ with $D_1=\frac{1-V}{2-\mu/t}$, where $V$ is the visibility in the interferometric detector, which can be established experimentally during the protocol setup, and $$H(\text{QBER})=-\log_2(\text{QBER})\text{QBER}-\log_2(1-\text{QBER})(1-\text{QBER})$$ is the binary Shannon entropy for the QBER.

In our model, we collect the noise photons coming from nonlinear mixing of the classical signals into an effective dark count rate 
\begin{equation}
p_d'=N_{F}+\gamma_{nl}\sum_{i} P_{tx,i}\cdot  \left(e^{-\alpha_c L_{i}}-e^{-\alpha_q L_{i}}  \right),
\end{equation} 
that will include a fixed number of photons from dark counts $N_F=p_d$ and the photons produced by nonlinear processes for each possible route between origin and destination. The constant $\gamma_{nl}$ absorbs all the fiber parameters and the detector efficiency $\eta$ so that we can get the expected number of noise photon detections in a single detection time bin. 

This corrected formula gives the number of ``false'' detections that are not triggered by a photon from Alice. As in other QKD security proofs, the rate formula we use assumes all the imperfections are caused by the presence of an eavesdropper. 

All the other parameters being equal, the QSNR decreases as the signal photons in the quantum channel attenuate during propagation. As soon as the losses and other imperfections in the channel reach a critical level, we can no longer guarantee we can extract a private key that cannot be guessed, totally or partially, by an attacker. This gives a distance limit for secure key generation.

	\section{Network Model}
	\label{Model}
	We can give a formal description of the QKD networks under study with a few reasonable assumptions. First, we will describe the exact problem we want to solve and then we will choose a relevant scenario to perform our numerical experiments.
	
	\subsection{Problem Statement}
	\label{Statement}
	We consider an arbitrary physical topology of an optical network as a directed graph $G(N,L, W_T,W_Q)$ composed of $N$ nodes and $L$ optical fiber links. $W_T$ is the total number of wavelengths in each link, while $W_Q$ is the number of reserved wavelengths for the quantum channel. We consider quantum requests $R(s,d)$ where $s$ is the source node and $d$ is the destination node. In the experiments, four channels are required to be established for each quantum request: the quantum channel, two classical control channels (one from $s$ to $d$ and one from $d$ to $s$), and an additional classical channel for user data transmission. 
	
	We want to provide a valid lightpath for all the traffic requests for the given topology. In our hybrid networks, the signals in the classical channels can disrupt the quantum transmission. We will propose RWA heuristics using power control over the classical channels to minimize the photons the classical signals generate into the quantum lightpaths and compare them to the usual classical heuristics.
	
	\subsection{Network Assumptions}
	We consider optical networks with neither wavelength converters nor amplifiers. To serve a connection request between any pair of nodes, not necessarily adjacent in the network, we establish a lightpath to which a route and a wavelength are allocated. Lightpaths whose routes share at least one link cannot be allocated the same wavelength. 
	
	We assume a QKD software-defined network (QKD SDN) with a classical control plane which uses the proposed algorithms to evaluate and allocate network resources for each connection request.
	
	For our experiments, we generate random geometric graphs corresponding to spatial networks on an Euclidean plane. We generate network topologies with differente numbers of nodes using two different graph models: Gabriel \cite{GS69} and Waxman \cite{Wax88}, both of which capturing some of the properties of typical communication networks. Each generated graph is constructed such that every edge becomes a bidirectional connection between two nodes, represented by two unidirectional links (i.e. one link from node $a$ to node $b$ and another from $b$ to $a$). Due to restrictions in the achievable secure distance for commercial QKD links, we have set relatively short links of tens of kilometers, ensuring that the longest path in the network does not exceed 60 km. To efficiently estimate the network diameter under these constraints—especially when computing exact longest paths is computationally expensive—we approximate it by selecting the longest path among the $k$-shortest paths between all node pairs.
	
	We assume static traffic, i.e., the connection requests are known in advance, although the control plane evaluates each request sequentially. The RWA algorithm considers the state of the network and uses the available resources to establish lightpaths. We consider two types of requests:
	
	\begin{itemize}
		\item {QKD channel demands.} This request requires establishing a quantum channel between a source and a destination node to generate a common secret key. The request will be associated to four lightpaths: a quantum channel to send the quantum states, two classical channels for bidirectional communication between the pair of nodes for synchronization and authenticated public discussions, and a classical channel to send the encrypted information with the generated key. Each channel requires a dedicated lightpath.
		
		\item{Classical channel demands.} We also consider the user can ask for purely classical connections on top of the QKD network.
	\end{itemize}

	The performance of the algorithms is measured in terms of {\bf blocking probability}, which is the ratio between the rejected connections and total demands. Connections for both quantum and classical requests are rejected if the algorithm cannot find a path and a wavelength between the source and destination nodes. Additionally, QKD lightpaths should be established only if they allow the generation of a secret key that meets a certain quality parameter, which we measure in terms of the Quantum Signal-to-Noise Ratio (QSNR). Furthermore, if any classical channel, either associated with the requested quantum channel or with a purely classical demand, causes noise to any established quantum channel and that noise introduces more errors than the protocol can tolerate, that connection should not be established, since it will cause a failure in the key generation procedure while wasting network resources. 
	
	Similarly, if the noise from classical existing signals would introduce into a quantum channel more errors than the protocol can tolerate, even if we have available wavelengths, the connection should not be established. This is because the key generation procedure will inevitably fail, leading to wasted network resources. To address this, we define a conservative threshold for the QSNR for each quantum lightpath, so that the connection is blocked if its QSNR falls below this threshold. 
	
	\section{Transmission Power Control and RWA Algorithms in QKD Networks}
	\label{ptx_control}
	
	\subsection{Transmission Power Control}
	We adapt the RWA algorithms described in \cite{RG23} by adding an adaptive power launch feature for the classical channels. The RWA algorithms find a candidate path and available wavelength and calculate the power launch for that particular path, so that the signal satisfies a given signal-to-noise ratio (SNR) at the receiver, following a similar procedure to \cite{BWB14, MBS21}. Once we have a route and know the fiber length $L_c$ the classical signal must travel, from the SNR expression in Equation (\ref{SNR}), we can calculate the launch power so that the SNR at the receiver matches the quality threshold in our network as
	\begin{equation}
		\label{eq:general_qsnr}
		P_{tx,\ c} = \frac{SNR\cdot N}{e^{-\alpha_{c} L_c}}.
	\end{equation}
	
	The control plane can use these calculations to adjust the power of the classical signal at each source transmitter to a level that both guarantees a clear reception and minimizes interference into the quantum channels (see Algorithm \ref{alg:launch_power}). 
	
		\begin{algorithm}
		\begin{small}
			\caption{Compute Launch Power to Meet SNR Threshold}
			\begin{algorithmic}[1]
				\State \textbf{global} $classicalAtt$
				\Procedure{ComputeLaunchPower}{$snrThreshold$, $ClassicalPath$}
				\State $totalAttenuation \gets 0$
				\State $length \gets 0$
				\For{$link \in ClassicalPath$}
				\State $length \gets length + link.length$
				\EndFor
				\State $totalAttenuation \gets e^{-classicalAtt \cdot length}$
				\State $launchPower \gets snrThreshold/totalAttenuation$
				\EndProcedure
			\end{algorithmic}
			\label{alg:launch_power}
		\end{small}
	\end{algorithm}

\subsection{RWA Algorithms}
	
	We study three different heuristics for routing the classical signals, both with and without power control. In the power control version, when we set a path, we set the power of the transmitter following Equation (\ref{eq:general_qsnr}).
	
	\begin{itemize}
		
		\item{Minimum Quantum Distance Overlap (MQDO):} This algorithm identifies all possible paths between a pair of nodes, marking the links of each path that are shared with established quantum channels. Then, the algorithm attempts to allocate an available wavelength on the path with the lower accumulated shared distance using the first-fit technique. If the allocation fails, the algorithm proceeds with the next path in ascending order of accumulated shared distance with established quantum channels and tries again. The process continues until finding a suitable wavelength or until all paths are exhausted. If the allocation is not possible, the connection is blocked. The pseudocode for this method is shown in Algorithm \ref{alg:mqdo}, which uses Algorithm \ref{alg:distance_overlap} as a subroutine. We also present here the procedure used in all the heuristics to check whether a new classical channel will reduce the QSNR of any existing quantum channel below the quality threshold (Algorithm \ref{alg:meetQSNR}) and the functions used in the calculation. Algorithm \ref{alg:QSNR} computes the QSNR of a quantum channel after we add a new classical channel using Equation (\ref{QSNR}), with Algorithm \ref{alg:RamanNoise} providing the Raman noise from the classical channels in the denominator. In the heuristics, whenever we have a request for a new classical connection, we will check if the QSNR of any existing quantum channel goes below a global quality threshold. If this is the case, that path is not selected. The connection request is blocked if there are no valid paths to establish the new classical channel.
		
		\begin{algorithm}
			\begin{small}
				\caption{Minimum Quantum Distance Overlap (MQDO)}
				\label{alg:mqdo}
				\begin{algorithmic}[1]
					\Procedure{Minimum Quantum Distance Overlap}{$source$, $destination$, $numWavelengths$, $candidatePathList$, $channelStatus$, $allocatedQuantumPaths$, $edgeLengthMatrix$}
					
					\State $candidatePathListCopy \gets$ \Call{copy}{$candidatePathList$}
					\State \Call{sort}{$candidatePathListCopy$, Distance Overlap}
					\For{$path$ in $candidatePathListCopy$}
					\If{\Call{meetsQSNR}{$path$}}
					\State $w \gets 0$
					\While{$(w < numWavelengths)$}
					\State $available \gets \TRUE$
					\For{$link \in path$}
					\If{$channelStatus[link][w] = 1$}
					\State $available \gets \FALSE$
					\EndIf
					\EndFor
					\If{$available$}
					\State \textbf{return} $path, w$
					\Else
					\State $w \gets w + 1$
					\EndIf
					\EndWhile
					\EndIf
					\EndFor
					\State \textbf{return} $[], -1$
					\EndProcedure
				\end{algorithmic}
			\end{small}
		\end{algorithm}

		\begin{algorithm}
			\begin{small}
				\caption{Distance Overlap (used in MQDO)}
				\begin{algorithmic}[1]
                    
					\State \textbf{global} $edgeLength$
					\State \textbf{global} $graphNetworkEdges$
					\State \textbf{global} $allocatedQuantumPaths$
					\Procedure{Distance Overlap}{$path$}
					\State $length \gets 0$
					
					\For{$link \in path$}
					\For{$key$ in $allocatedQuantumPaths$}
					\For{$qPath$ in $allocatedQuantumPaths[key]$}
					\If{$link \in qPath$}
					\State $length \gets length + edgeLength[link]$
					\EndIf
					\EndFor
					\EndFor
					\EndFor
					\State \textbf{return} $length$
					\EndProcedure
				\end{algorithmic}
				\label{alg:distance_overlap}
			\end{small}
		\end{algorithm}

 \begin{algorithm}
            \begin{small}
                \caption{Meet QSNR}

                \begin{algorithmic}[1]
                    \State \textbf{global} $quantumChannels$
                    \State \textbf{global} $qsnrThres$
\Procedure{meetQSNR}{$classicalPath$}
                        \For{$quantumPath \in quantumChannels$}:
                        \If{$\Call{getQsnr}{quantumPath,classicalPath} < qsnrThres$}
                            \State \textbf{return} $False$
                        \EndIf
                    \EndFor
                    \State \textbf{return} $True$

				\EndProcedure
                \end{algorithmic}
			\label{alg:meetQSNR}
         \end{small}
        \end{algorithm}

 \begin{algorithm}
            \begin{small}
              \caption{QSNR calculation}
				\label{alg:QSNR}
                \begin{algorithmic}[1]

					\State \textbf{global} $graphNetworkEdges$
                    \State \textbf{global} $edgeLength$
                    \State \textbf{global} $quantumAtt$
                    \State \textbf{global} $eta$ 
                    \State \textbf{global} $fixedNoise$
                 \Function{getQSNR}{$quantumPath,NewClassicalPath$}
                
                \State $length \gets 0$
                \ForAll{$link \in quantumPath$}
                    \State $s \gets graphNetworkEdges[link][0]$
                    \State $d \gets graphNetworkEdges[link][1]$
                    \State $length \gets length + edgeLength[s][d]$
                \EndFor
                \State $transmitivity \gets \exp(- quantumAtt  \cdot length)$
                \State $launchPower \gets eta \cdot transmitivity$

                \State $sharedNoise \gets
\Call{CalculateNoise}{quantumPath, NewClassicalPath}$
                \State $result \gets \frac{transmitivity \cdot launchPower}{fixedNoise + sharedNoise}$

                \State \Return $result$
                \EndFunction

            \end{algorithmic}
            \end{small}
        \end{algorithm}

        \begin{algorithm}
            \caption{Calculate Noise due to Classical Channels}
				\label{alg:RamanNoise}
            \begin{algorithmic}[1]
        \State \textbf{global} $quantumChannels$ 
		\State \textbf{global} $classicalChannels$      
		\State \textbf{global} $graphNetworkEdges$ 
		\State \textbf{global} $edgeLength$
		\State \textbf{global} $classicalAtt$
		\State \textbf{global} $quantumAtt$
	    \State \textbf{global} $nonlinearParam$

\Function{CalculateNoise}{$quantumPath,newClassicalPath$}
                \State $RamanNoise \gets 0$
                
                \ForAll{$classicalPath \in classicalChannels$} \hfill \Comment{Existing channels} 
                        \For{$i \gets 0$ \textbf{to} $length(classicalPath) - 1$}
                            \State $link \gets classicalPath[i]$
                            \If{$link \in quantumPath$}
                                \State $(s, d) \gets graphNetworkEdges[link]$
                                \State $linkLength \gets edgeLength[s][d]$
                                \State $power \gets classicalPath.LinkPower[i]$
                                \State $RamanNoise \gets RamanNoise + power
\cdot \left(e^{-classicalAtt \cdot linkLength} - e^{-quantumAtt \cdot linkLength}\right)$
                            \EndIf

                    \EndFor                                   
                \EndFor

                  \For{$i \gets 0$ \textbf{to} $length(newClassicalPath) - 1$} \hfill \Comment{New channels} 
                            \State $link \gets newClassicalPath[i]$
                            \If{$link \in quantumPath$}
                                \State $(s, d) \gets graphNetworkEdges[link]$
                                \State $linkLength \gets edgeLength[s][d]$
                                \State $power \gets newClassicalPath.LinkPower[i]$
                                \State $RamanNoise \gets RamanNoise + power \cdot \left(e^{-classicalAtt \cdot linkLength} - e^{-quantumAtt \cdot linkLength}\right)$
                            \EndIf

                    \EndFor

                \State \Return $nonlinearParam \cdot RamanNoise$
                \EndFunction
            \end{algorithmic}
        \end{algorithm}

		\item{Minimum Quantum Classical Channel Overlap (MQCCO):} Similar to MQDO, the algorithm calculates the accumulated shared distance with quantum connections for each possible path, considering other established classical channels sharing any link of the path. Specifically, for each established quantum channel, if a link of the candidate path is marked as shared, and that link is also traversed by an established classical channel, the contribution to the accumulated shared distance is doubled. After that, the algorithm proceeds as described in MQDO. It tries to allocate an available wavelength to a candidate path using the first-fit technique. If the allocation is not possible, the algorithm moves to the next path in increasing order of shared distance. If the algorithm is unable to allocate resources after evaluating all candidate paths, the request is blocked. Algorithm \ref{alg:mqcco} shows the pseudocode for MQCCO, which uses Algorithm \ref{alg:channel_overlap} as a subroutine.

		\begin{algorithm}
			\begin{small}
				\caption{Minimum Quantum-Classical Channel Overlap (MQCCO)}
				\begin{algorithmic}[1]
					\Procedure{Minimum Quantum-Classical Channel Overlap}{$source$, $destination$, $numWavelengths$, $candidatePathList$, $channelStatus$, $allocatedQuantumPaths$}
					
					\State $sortedCandidatePathList \gets$ \Call{copy}{$candidatePathList$}
					\State \Call{sort}{$sortedCandidatePathList$, Channel Overlap}
					\For{$path$ in $sortedCandidatePathList$}
					\If{\Call{meetsQSNR}{$path$}}
					\State $w \gets 0$
					\While{$(w < numWavelengths)$}
					\State $available \gets \TRUE$
					\For{$link \in path$}
					\If{$channelStatus[link][w] = 1$}
					\State $available \gets \FALSE $
					\EndIf
					\EndFor
					\If{$available$}
					\State \textbf{return} $path, w$
					\Else
					\State $w \gets w + 1$
					\EndIf
					\EndWhile
					\EndIf
					\EndFor
					\State \textbf{return} $[], -1$
					\EndProcedure
				\end{algorithmic}
				\label{alg:mqcco}
			\end{small}
		\end{algorithm}

		\begin{algorithm}
			\begin{small}
				\caption{Channel Overlap (used in MQCCO)}
				\begin{algorithmic}[1]

					\State \textbf{global} $edgeLength$
					\State \textbf{global} $graphNetworkEdges$
					\State \textbf{global} $allocatedQuantumPaths$
					\State \textbf{global} $channelStatus$
					\Procedure{Channel Overlap}{$path$}
					\State $length \gets 0$
					
					\For{$link$ in $path$}
					\State $occupiedWavelengths \gets$ \Call{sum}{$channelStatus[link]$}
					\For{$key$ in $allocatedQuantumPaths$}
					\For{$qPath$ in $allocatedQuantumPaths[key]$}
					\If{$link \in qPath$}
					\State $length \gets length + occupiedWavelengths \times edgeLength[link]$
					\EndIf
					\EndFor
					\EndFor
					\EndFor
					\State \textbf{return} $length$
					\EndProcedure
				\end{algorithmic}
				\label{alg:channel_overlap}
			\end{small}
		\end{algorithm}

		\item{Quantum Totally Disjoint (QTD):} The algorithm attempts to allocate a path and an available wavelength that do not share links with any established quantum channel. If the algorithm cannot find such a path or available spectral resources, the connection is blocked. The pseudocode is shown in Algorithm \ref{alg:qtd}. 
	\end{itemize}

	\begin{algorithm}
		\begin{small}
			\caption{Quantum Totally Disjoint (QTD)}
			\begin{algorithmic}[1]
				\Procedure{Quantum Totally Disjoint}{$source$, $destination$, $numWavelengths$, $candidatePathList$, $channelStatus$, $allocatedQuantumPaths$}
				\For{$path$ in $candidatePathList$}
				\If{\Call{meetsQSNR}{$path$}}
				\State $search \gets \TRUE $
				\State $foundPath \gets \TRUE $ 
				\While{$search$}
				\For{$key$ in $allocatedQuantumPaths.keys()$}
				\For{$qPath$ in $quantum\_paths[key]$}
				\For{$link \in path$}
				\If{$link \in qPath$}
				\State $search \gets \FALSE$
				\State $foundPath \gets \FALSE $
				\State \textbf{break}
				\EndIf
				\EndFor
				\EndFor
				\EndFor
				\State $search \gets \FALSE$
				\EndWhile
				\EndIf
				\If{$foundPath$}
				\State $w \gets 0$
				\While{$(w < numWavelengths)$}
				\State $available \gets \TRUE$ 
				\For{$link \in path$}
				\If{$channelStatus[link][w] = 1$}
				\State $available \gets \FALSE $
				\EndIf
				\EndFor
				\If{$available$}
				\State \textbf{return} $path, w$
				\Else
				\State $w \gets w + 1$
				\EndIf
				\EndWhile
				\EndIf
				\EndFor
				\State \textbf{return} $[], -1$
				\EndProcedure
			\end{algorithmic}
			\label{alg:qtd}
		\end{small}
	\end{algorithm}

	All these proposals are compared to the reference classical RWA algorithm $k$-Shortest Paths (KSP) and First Fit (FF) \cite{ZJM00} for both the quantum and the classical channels, integrating the transmission power control protocol for the classical channels. The algorithm computes the $k$ shortest paths between the source and destination. Subsequently, in increasing order of path length, the algorithm attempts to allocate the first available wavelength along each path until a suitable pair of path-wavelength is found. If the algorithm cannot allocate the required resources, the request is blocked.
	
	\subsection{Quantum channel evaluation}
	\label{Qchan}
	Quantum channels are assigned using a $k$-shortest-path-first-fit heuristic. Losses are higher in the O band and we prefer the paths with shorter lengths (higher transmission). For the quantum channel, all the proposed algorithms verify if the selected path meets the requirements of QSNR. In the case of quantum requests, if the resulting QSNR falls below a certain threshold, the algorithms sequentially evaluate the next path until they identify a path with a free wavelength that satisfies the QSNR threshold or block the request otherwise. 
	
	Conversely, the establishment of classical channels must not degrade the performance of existing quantum connections. Hence, the RWA algorithms ensure that the selected path does not take the QSNR of established quantum channels below the specified QSNR threshold. If a path fails to meet this criterion, the algorithms proceed to assess subsequent paths until finding one that does not degrade the established quantum channels. If no such path is found, the connection is blocked. We estimate the value of the QSNR with Algorithm \ref{alg:QSNR}, which follows Equation (\ref{QSNR}).
	
	Finally, for the QTD algorithms we enforce separate classical and quantum paths when we assign the quantum channels. Any path sharing a single link with a classical signal leads to a blocked request.
	
	\section{Simulation Set up and Results}
	\label{Results} 
	\subsection{Set up}
\begin{figure}
		\centering
		
		\includegraphics[width=0.45\linewidth]{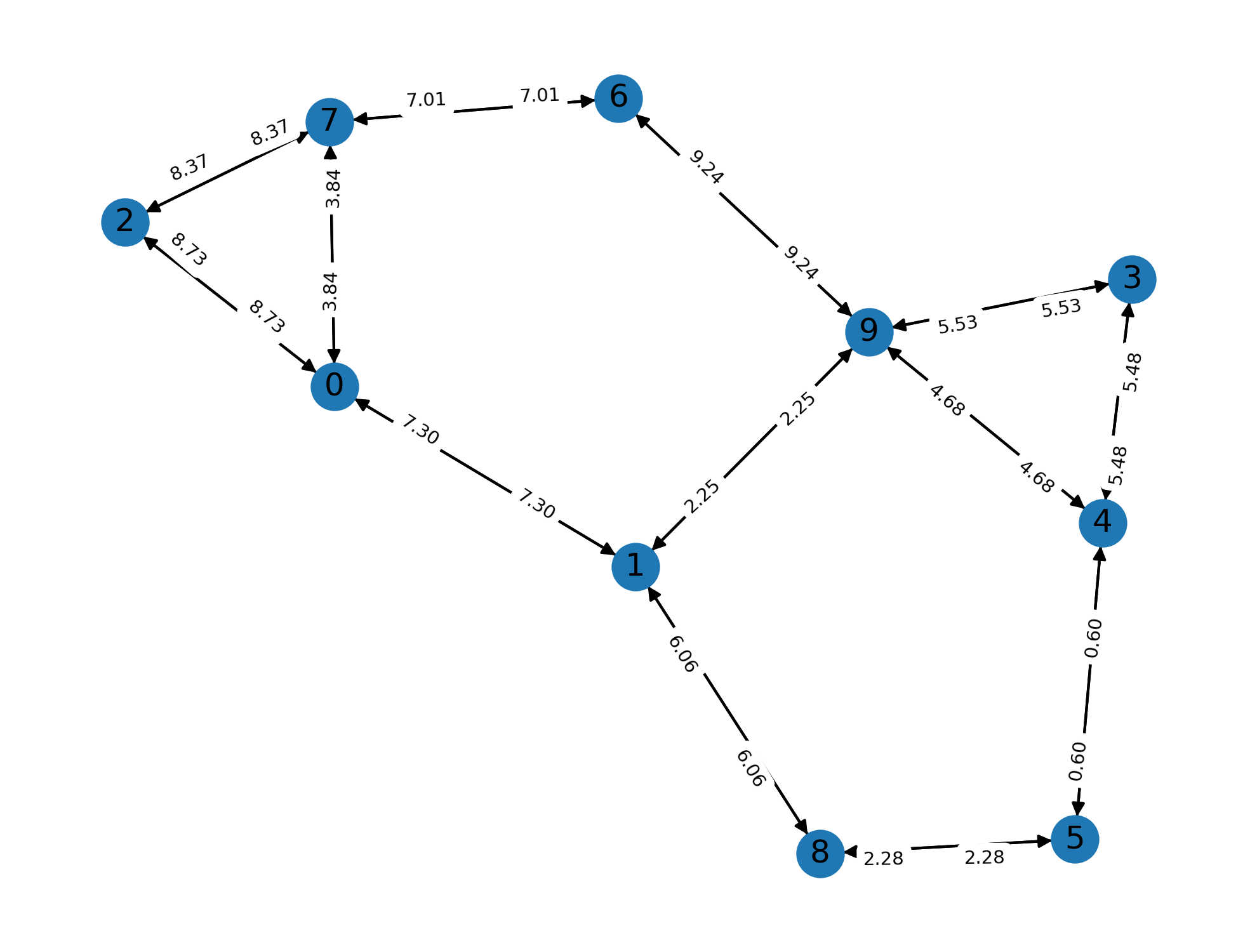}      \includegraphics[width=0.45\linewidth]{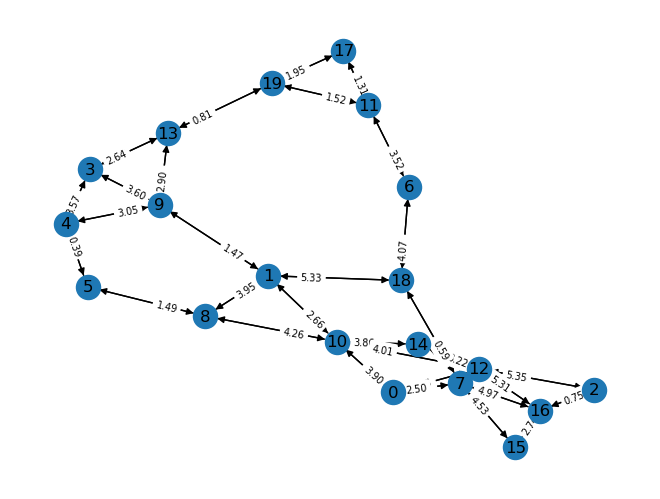}\\	
        \includegraphics[width=0.45\linewidth]{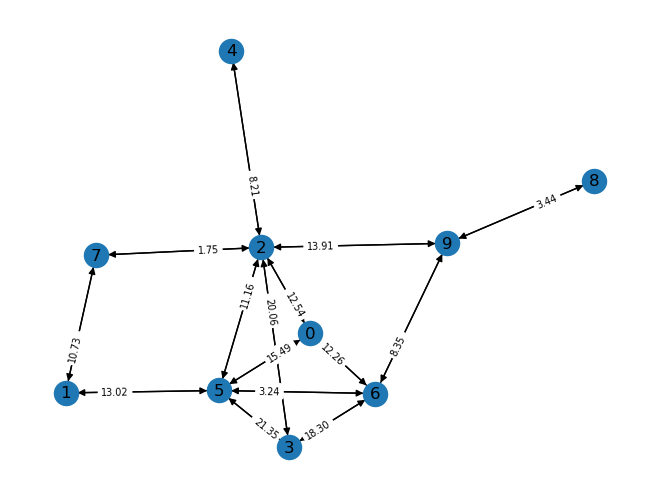}   \includegraphics[width=0.45\linewidth]{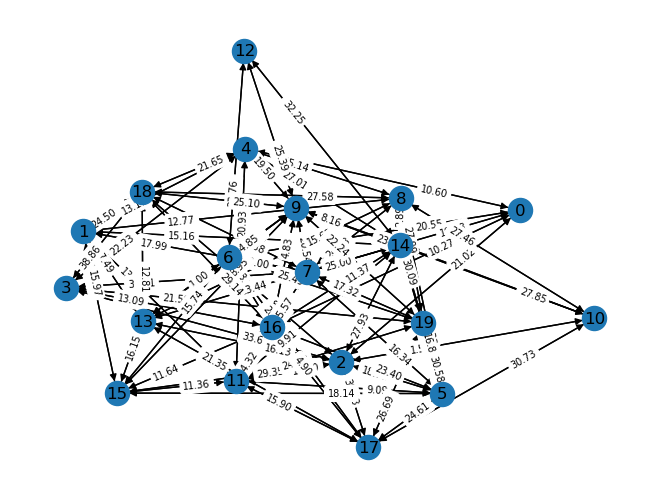}
		\caption{Example of four random topologies generated in our simulations using Gabriel (top) and Waxman (bottom) graphs with $10$ (left) and $20$ nodes (right).}\label{fig:random_top_0}
		
	\end{figure}

	We investigate the effects of the transmission power protocols on $100$ randomly generated topologies for each graph generation model. Each topology is generated with either $10$ or $20$ nodes where each node is connected to at least one neighbour. Waxman graphs are generated with parameters $\alpha=100$ and $\beta=0.45$, adding links, if necessary, to make sure the graph is connected (there are no orphan nodes or disconnected subgraphs). Examples of the generated topologies are illustrated in Figure \ref{fig:random_top_0}.
	
	We assume $50$ wavelengths per fiber, $10$ for the quantum channels, in the O-band, which generally implements Coarse WDM with a wider channel spacing (around 20 nm) \cite{G.694.2}, and $40$ for the classical channels in the C-band, assuming Dense WDM with a 100 GHz channel separation \cite{G.694.1}. The noise terms $N_f$ and $\gamma_{nl}$ are set to $10^{-4}$ and $2.25\cdot 10^{-4}$, respectively, considering dark counts dominate the fixed noise and the Raman noise model described in Section \ref{LinkModel}. We consider a normalized $P_{tx,q}$ where 1 is the power of a single photon per pulse and two fiber attenuation coefficients: $\alpha_{q}=\SI{0.32}{\deci\bel/\kilo\meter}$ around \SI{1310}{\nano\meter} and $\alpha_{c}= \SI{0.17}{\deci\bel/\kilo\meter}$ around \SI{1550}{\nano\meter}, each corresponding to standard silica optical fiber at the usual wavelengths in the O and C bands assigned to the quantum and classical channels, respectively, in many QKD hybrid networks. Channels can be established only if a candidate path with available resources is found and the QSNR of the quantum lightpath at reception is, at least, \SI{-5}{\deci\bel}.
	
	We also normalize the launch power of the classical channels with respect to the maximum launch power. This maximum power launch is determined by ensuring that the longest possible path for that particular connection achieves a minimum power of \SI{-36.8}{\dBm} at the receiver. For realistic detectors and noise levels, this power has shown to be enough for \SI{1.25}{\giga\bit/\second} data rates in a fiber network without amplifiers \cite{PDC12}. For power control, the RWA algorithms identify the candidate path and they compute the required transmission power for that particular path and normalize it to the maximum launch power. Consequently, both $P_{tx,q}$ and $P_{tx,c}$ are normalized at their respective scales. Since the transmission power of quantum and classical lightpaths differ in magnitude, this difference is accounted for in the $\gamma_{nl}$ term and in the calculations giving the QSRN and the key rate. 
	
	Lastly, the detector quantum efficiency is set to $\eta=0.1$, and the visibility of the interferometer at the receiver is assumed to be $V=0.98$. For these parameters, and considering a BB84 protocol, we set $P_{tx,q}=\eta t$, which takes into account the effects of imperfect detection and assumes we have a mean photon number at the transmitter equal to the transmissivity of the quantum channel $t<1$, chosen to optimize the final key rate \cite{NSG05}.
	
	We use the request blocking ratio and the average SNR as the figures of merit to evaluate the performance of the algorithms. The total number of requests arriving to the network varies from $10$ to $100$ with an increment of $10$ requests. For each topology, $50$ simulations are run. Figures include $95\%$ confidence intervals. The averages for different runs of the experiments are fairly stable and the confidence intervals are difficult to appreciate in some of the Figures for the relevant scale. In some case, variability is reduced with a higher network traffic (see, for instance, Figure \ref{fig:blocking}).

	\subsection{Results}
	\label{sub:results}

	\begin{figure}
		\centering
		
		\includegraphics[width=0.45\linewidth]{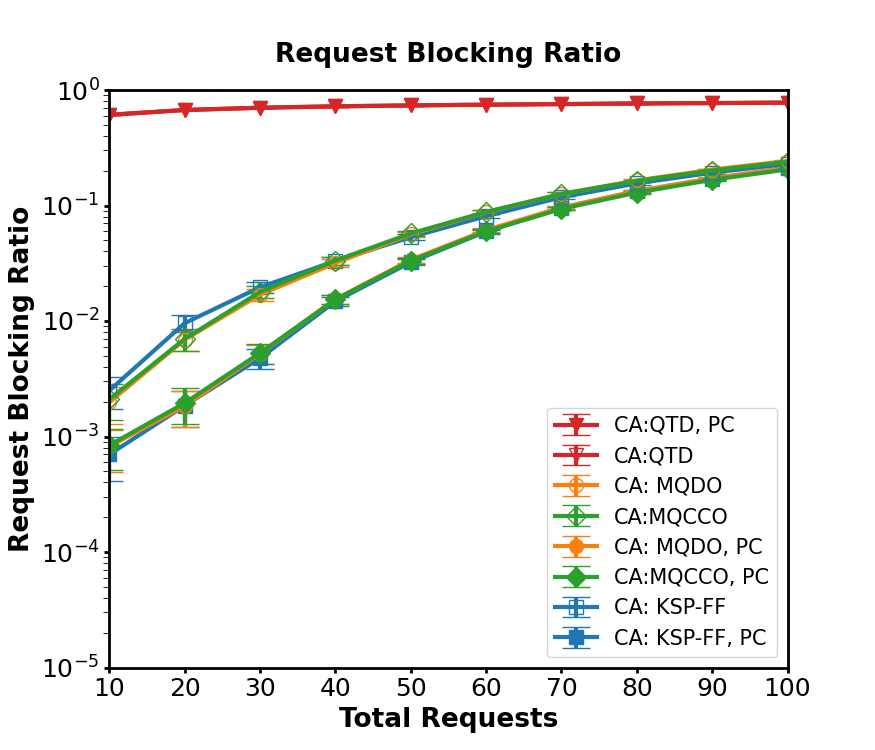}     
		\caption{Blocking ratios obtained by the RWA algorithms with and without power control.}\label{fig:blocking}
		
	\end{figure}
	
	\begin{figure}
		\centering
		
		\includegraphics[width=0.45\linewidth]{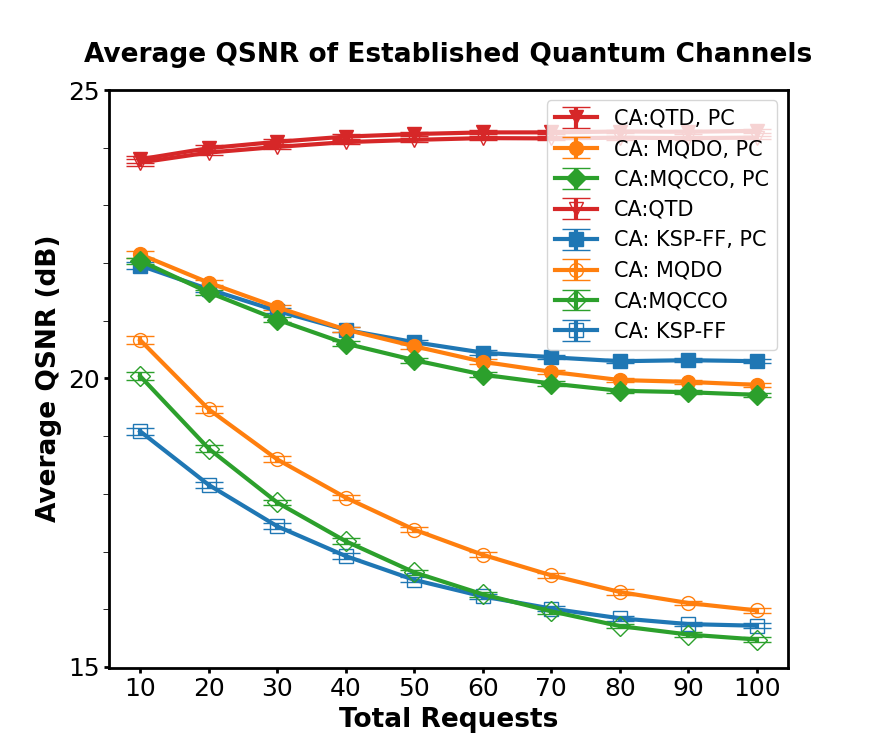}     \includegraphics[width=0.45\linewidth]{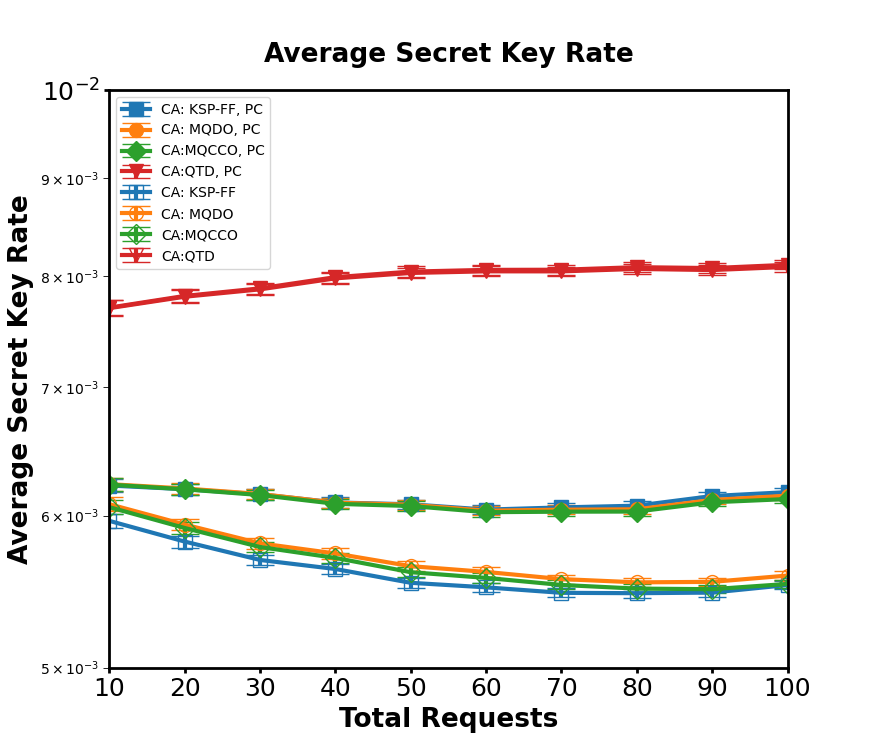}
		\caption{QSNR of the established QKD channels using the proposed algorithms (left) and average Secret Key Rate in bits per pulse (right).}\label{fig:quality_figures_10}
			
		\end{figure}
		
	Figure \ref{fig:blocking} shows the blocking ratio of the algorithms with transmission power control in Gabriel topologies with 10 nodes. The blocking ratio is reduced by up to an order of magnitude for the RWA algorithms that implement the power control protocol. This result demonstrates that reducing the power control of the classical channels helps to meet the QSNR requirements, thereby decreasing the blocking ratio and enhancing the performance of the QKD network. This improvement is not observed for the QTD algorithms, with blocking ratios above $10^{-1}$, since they are designed to find separate routes for QKD and classical channels. Therefore, the blocking ratio is primarily due to lack of available resources for establishing new channels, rather than failing to meet the QSNR requirements as in the other algorithms. 
		
	Figure \ref{fig:quality_figures_10} shows the achieved QSNR (left) and the average Secret Key Rate, SKR, (right) for the quantum channels that are established. The average QSNR is consistently above \SI{15}{\deci\bel} across all algorithms. Without active transmission power control, the QSNR decreases, as expected, due to increased noise on the quantum lightpaths caused by the higher transmission power of the classical channels. A smaller QSNR is related to a higher QBER which, in turn, means the QKD protocol has a slower key generation rate.

In our results, all the Figures show the final key bit rate per pulse following the estimation in Equation (\ref{eq:eqRBB84}), which can be related to the total secret key rate in bits per second \cite{SBC09}. We have considered a system with detector efficiency $\eta=0.1$ and attenuated coherent pulses with an average photon number $\mu=t$, where $t$ is the transmissivity of the quantum channel and has been chosen to optimize the final key rate \cite{NSG05}. In that case, the expected number of photon detections per pulse, considering a maximum distance of \SI{60}{\kilo\meter} and our attenuation constant in the O band, is $n=\eta t^2=\eta 10^{-0.2\alpha_{qdB}}\approx 1.45\cdot 10^{-4}$. Typical QKD lasers can reach pulse repetition rates around \SI{10}{\giga\hertz}, which would give a photon detection rate of, at least, \SI{1.45}{\mega\hertz}. If we consider this is the case and that the limiting factor for protocol detections is the detector dead time, which we can set in the microsecond range, the detection rate can be around \SI{1}{\mega\hertz}. For these settings, the final secret key rate in our simulations would be in the range of a few kilobits per second. 

	Regarding the performance of each heuristic, KSP-FF achieves a QSNR between 16-\SI{19}{\deci\bel} without power control, with an improvement of up to \SI{5}{\deci\bel} observed for higher number of requests when power control is enabled. Both MQDO and MQCCO exhibit similar trends; MQDO without power control delivers a QSNR advantage of \SI{1}{\deci\bel} over MQCCO and \SI{2}{\deci\bel} over SP-FF at low request volumnes. For both MQCCO and MQDO, the QSNR improves up to $5$ dB with power control.
	
	When power control is inactive, all of the algorithms yield lower quality channels compared to quantum-oriented RWA algorithms. The lower quality of the quantum channels translates into slower key generation rate as shown in Figure \ref{fig:quality_figures_10} (right). The algorithms without power control achieve lower average key generation rates compared to the algorithms with power control. In this case, the QTD algorithm achieves the fastest generation rate since the interference caused by the classical channels in the QKD channels is the lowest among the proposed algorithms, but at the cost of a higher blocking probability.
	
		\begin{figure}
		\centering
		
		\includegraphics[width=0.45\textwidth]{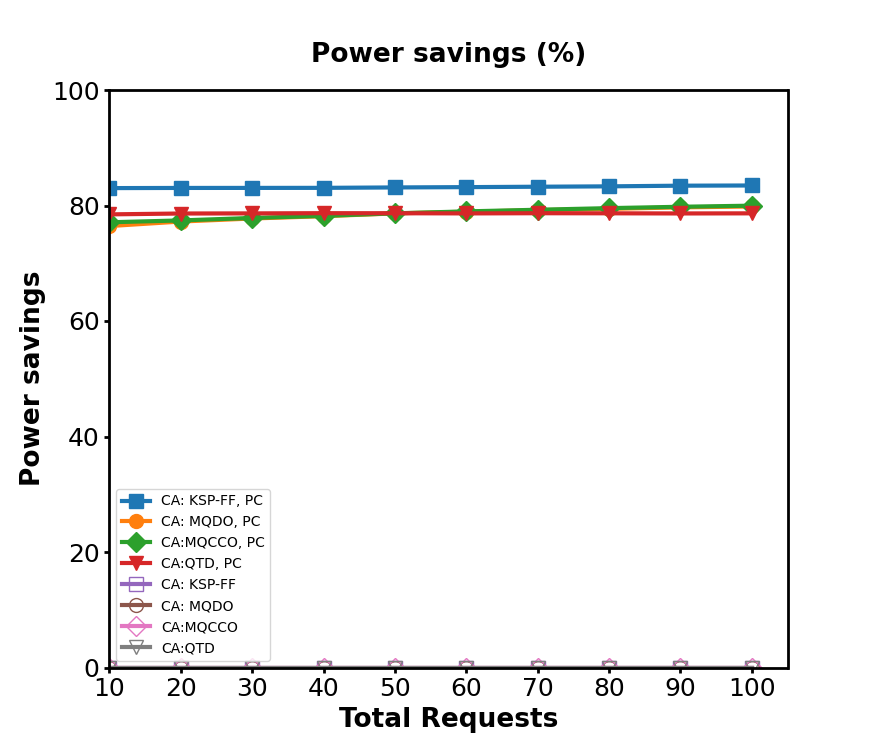}     
		\caption{Percentage of power consumption reduction for the RWA algorithms with enabled power control}\label{fig:power_savings_10}
		
	\end{figure}
	Lastly, enabling a power control protocol can save up to $80\%$ of energy for all algorithms, as shown in Figure \ref{fig:power_savings_10}. Power consumption is calculated from the mean launch power of the established channels. Subsequently, the power savings are calculated using the following formula:
\begin{equation}
     \frac{P_{tx, no PC}-P_{tx, PC}}{P_{tx, no PC}}.
\end{equation}
This expression quantifies the reduction in the average launch power consumption of the established channels when the power control feature is active $P_{tx, PC}$, relative to the consumption when the feature is inactive, $P_{tx, no PC}$. The greatest savings happen for shortest path firtst fit with power control. This is logical, as shorter paths mean there are less losses and a larger excess power with respect to the threshold at the detector. The quantum-oriented heuristics choosing longer paths to avoid contamination into the quantum signals have a slightly lower power reduction.
	
	These experiments suggest proper control of classical transmission power can enhance signal quality and enables the use of classical RWA algorithms for solving the resource allocation problem, while reducing operational expenditures due to the reduction of power consumption. 
	
	\begin{figure}
		\centering
		
		\includegraphics[width=0.45\linewidth]{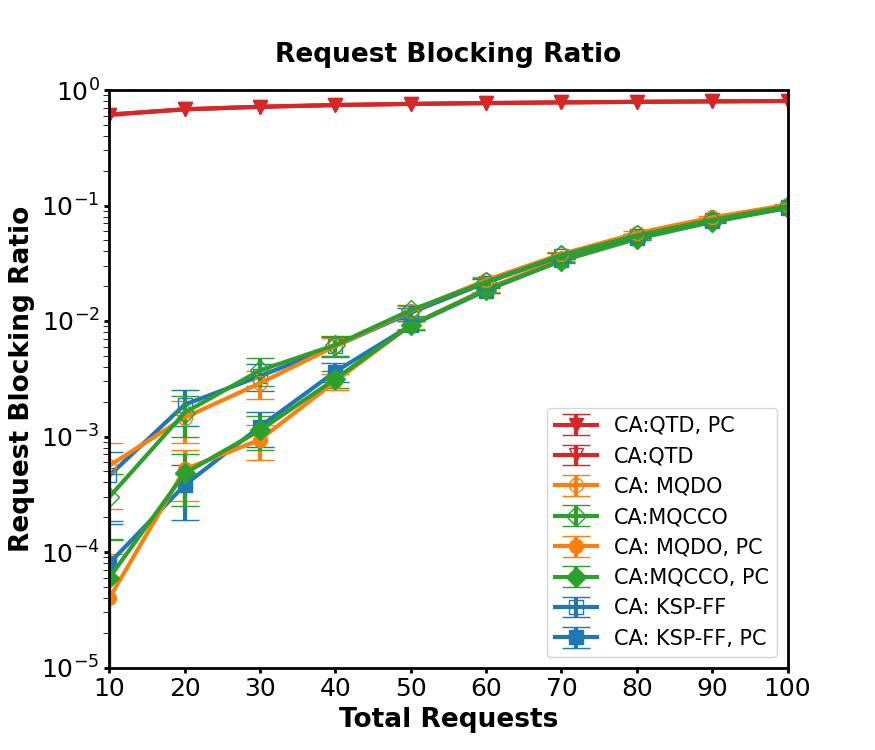}     \includegraphics[width=0.45\linewidth]{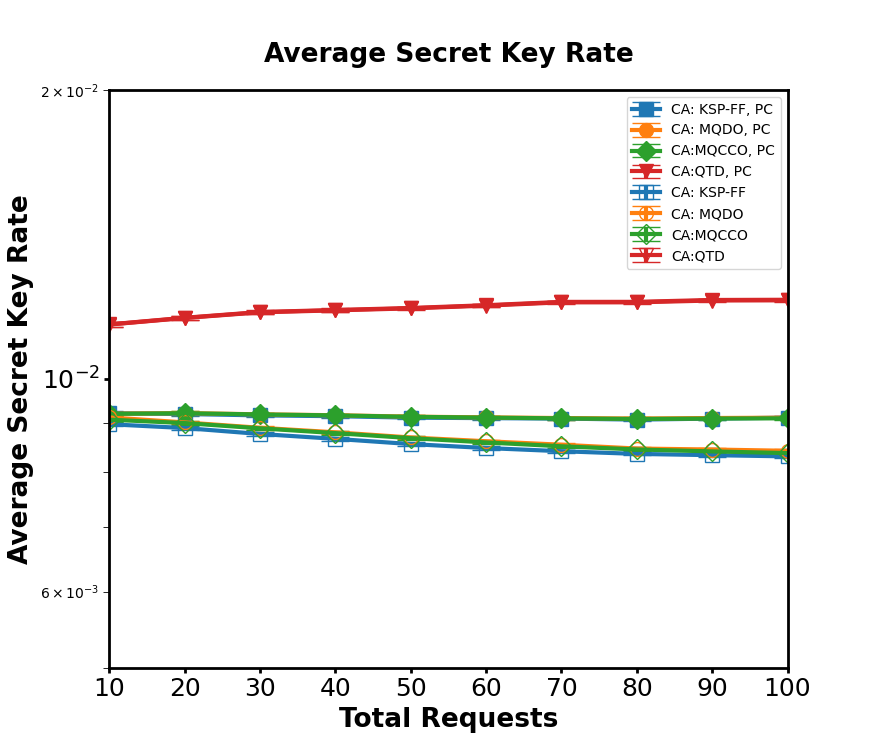}
		\caption{Blocking ratio and SKR for random Gabriel topologies of $20$ nodes.}\label{fig:blocking_20}
		
	\end{figure}

	Results present a similar trend when the number of nodes is increased to $20$. The blocking ratio is lower, as shown in Figure \ref{fig:blocking_20} (right). Since the network topologies are larger while the traffic volume remains unchanged, the blocking ratio decreases due to the increased availability of resources and the greater number of possibilities for accommodating classical channels without interfering with the QKD channels. On the other hand, RWA algorithms with power control achieve a better blocking ratio, however, the improvement is less pronounced compared to the $10$-node topologies, likely because the blocking ratios are already significantly lower. Having a larger network allows for better alternative routes and less connection requests are rejected. In this scenario with a larger network, we also see a higher variance and the confidence intervals in the simulations are quite visible in the blocking probability, particularly for a smaller number of requests, when the particular topology of the random network is more important.

The SKR also improves with respect to the $10$-node topologies, obtaining key rates of up to $9\cdot 10^{-3}$ bits/pulse for SP-FF, MQDO and MQCCO and higher than $1\cdot 10^{-2}$ for QTD. Also, results show that when the traffic volume increases, RWA algorithms without power control produce slower key rates. Hence, QKD networks must have low traffic requirements in order to maintain high-quality QKD connections.

\begin{figure}
		\centering
		
		\includegraphics[width=0.45\linewidth]{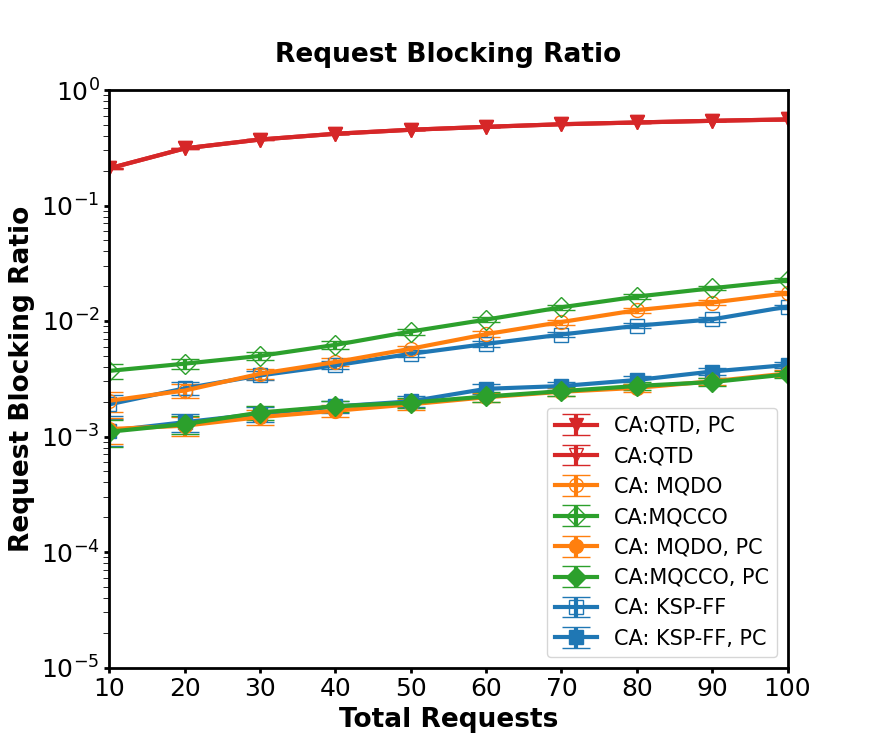}     
		\includegraphics[width=0.45\linewidth]{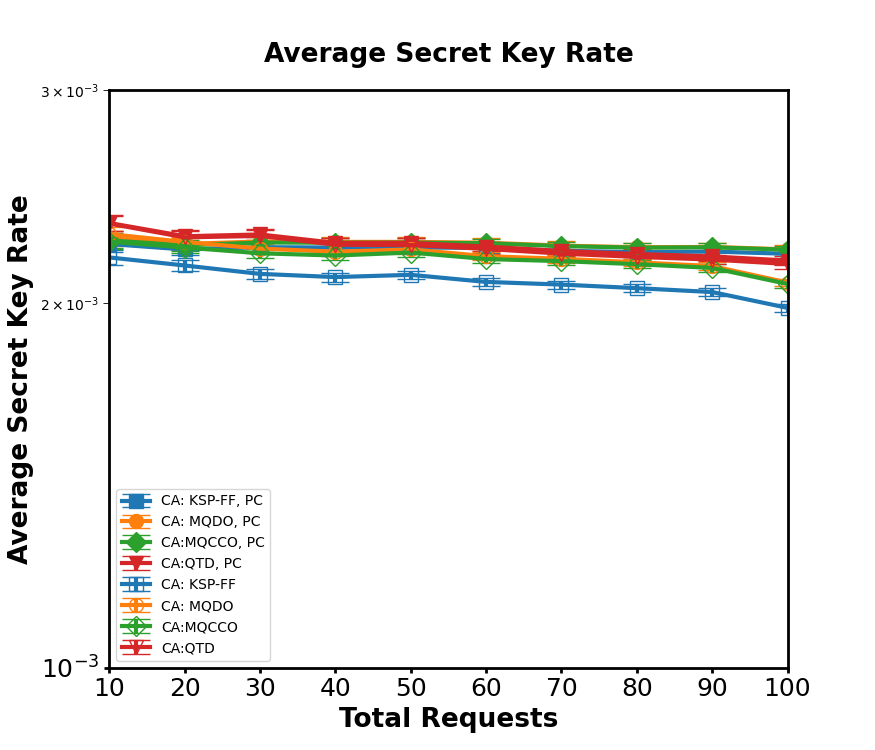}
		\caption{Blocking ratio and SKR for random Waxman topologies of $20$ nodes.}\label{fig:br_waxman}
		
	\end{figure}

	Lastly, we present the results for $20$-node random Waxman topologies. The chosen $\alpha$ and $\beta$ parameters of the Waxman distribution have been selected to create networks with a medium density of connections. In this case, the blocking probability increases with respect to the blocking probability in the $20$-node Gabriel topologies, as shown in Figure \ref{fig:br_waxman} (left). This result suggests that the generated Waxman networks have more difficulties for accomodating the channels and meeting the QSNR requirements. For this reason the key generation rate is also lower compared to the Gabriel graphs, as shown in Figure \ref{fig:br_waxman} (right), where the key rates are slightly higher than  $2\cdot 10^{-3}$ bits/pulse. Interestingly, enabling power control does not result in a significant improvement in the key generation rate, which points to the limited availability of paths that meet the QSNR requirements as the key factor for the blocking ratio and the lack of quality in the QKD connections. However, power control does reduce the blocking ratio, up to an order of magnitude under high load conditions.

	\subsection{Performance of mixed traffic scenarios.}
	\label{sub:classical_traffic}

	We also investigate power control in RWA algorithms in mixed traffic scenarios, i.e., networks in which both quantum and classical requests are present. We consider a network trying to establish QKD links over a network with classical traffic outside our control. In order to do that, we study the effects of an increasing number of classical traffic demands. With these simulations, we try to understand how to balance network resources between classical and quantum channels.
	
	Specifically, we analyze the performance of the network for $90$ total requests, which gives a flexible scenario with enough classical channels to be interesting, but not too many to prevent efficient QKD. We define the classical traffic load as the fraction of the total requests that belong to purely classical traffic. We consider these classical channels are confined to our network and that the signals have never crossed an amplifier. Additionally, we suppose these additional channels are not under our control and they do not follow the rules of our heuristics. Their route is fixed, but we can attenuate them (add power control).
	
	The classical load varies from $0$ (representing all the request are quantum, including the request for classical service channels) to $1$ (representing all the requests are classical). All other simulation parameters remain unchanged and results are represented with $95\%$ confidence intervals. 
	
	\begin{figure}
		\centering
		\includegraphics[width=0.5\linewidth]{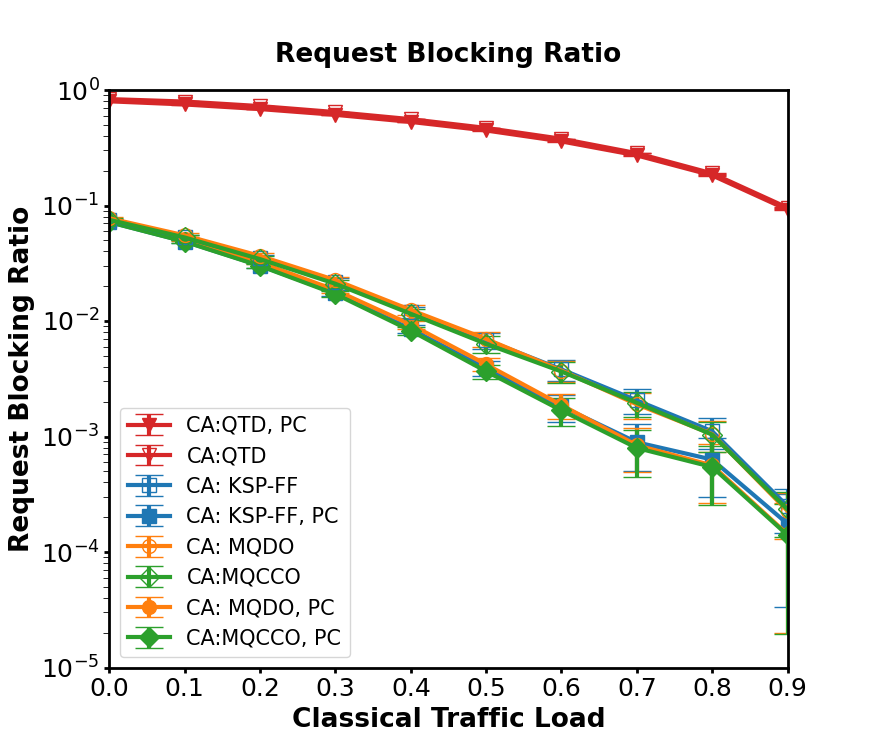}
		\caption{Blocking ratio in hybrid networks for $20$-node random Gabriel topologies.}
		\label{fig:br_classical}
	\end{figure}
	
	\begin{figure}
		\centering
		\includegraphics[width=0.45\linewidth]{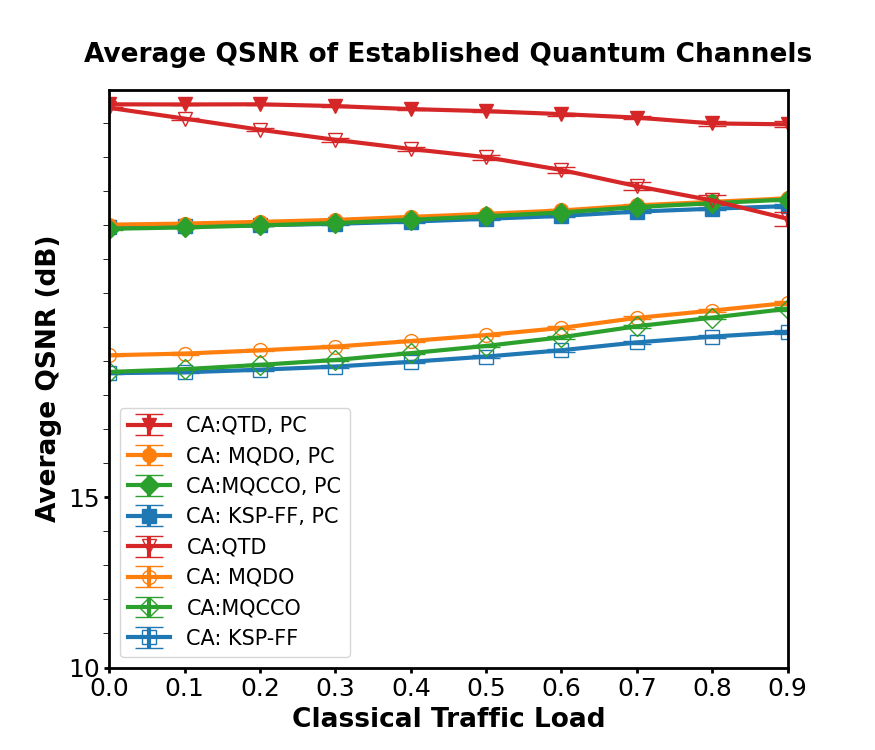}
		\includegraphics[width=0.45\linewidth]{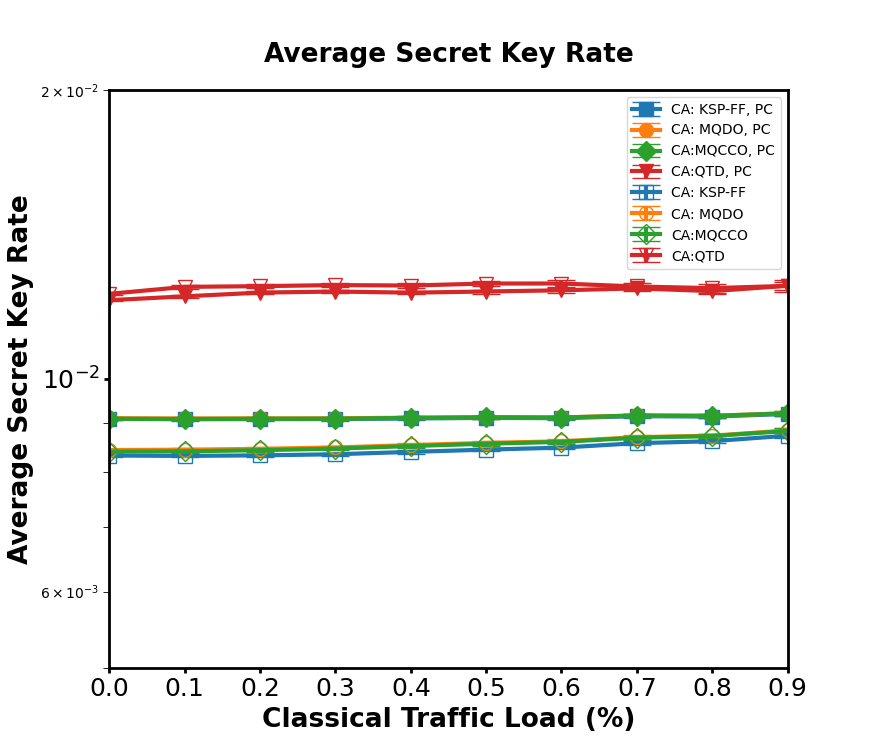}
		\caption{QSNR and SKR hybrid networks for $20$-node random Gabriel topologies.}
		\label{fig:snr_kr_classical}
	\end{figure}
	
	Figure \ref{fig:br_classical} shows the blocking ratio of the RWA algorithms with power control protocol and when this protocol is not active for $20$-node random Gabriel topologies. We only show the $20$-node results to avoid repetition, since the trends are similar for the $10$-node random Gabriel graphs and the Waxman graphs. 
	
	As in Figure \ref{fig:blocking}, introducing power control results in a lower blocking ratio, although the effect is not so significant, at least for the selected traffic load, as shown in Figure \ref{fig:br_classical}. Additionally, the blocking ratio also decreases when the classical traffic load increases. Since a classical connection only requires the establishment of one lightpath, the accommodation is easier compared to a quantum request, which requires the establishment of four different lightpaths. We can also see that, for a high fraction of classical traffic, the blocking ratio can be sensitive to the particular quantum connections required and the network topology, as shown by the increased variability evinced by the wider confidence intervals.
	
	In terms of QSNR, Figure~\ref{fig:snr_kr_classical} (left) confirms that enabling power control significantly enhances the quality of QKD connections, with improvements of up to \SI{6}{\deci\bel}. This translates into faster secret key generation, as illustrated in Figure~\ref{fig:snr_kr_classical} (right). In both cases, the best performance—characterized by the highest QSNR and the fastest key generation rate—is achieved by QTD with power control enabled. For the previous scenarios, QTD gave the same results with and without power control. The classical and quantum paths never shared the same fiber. Now, the classical traffic outside our direct control, which is not associated to QKD channels, can be routed on top of an existing quantum link. Adjusting the power of these channels reduces their noise. 
 
\begin{figure}
		\centering
		\includegraphics[width=0.5\linewidth]{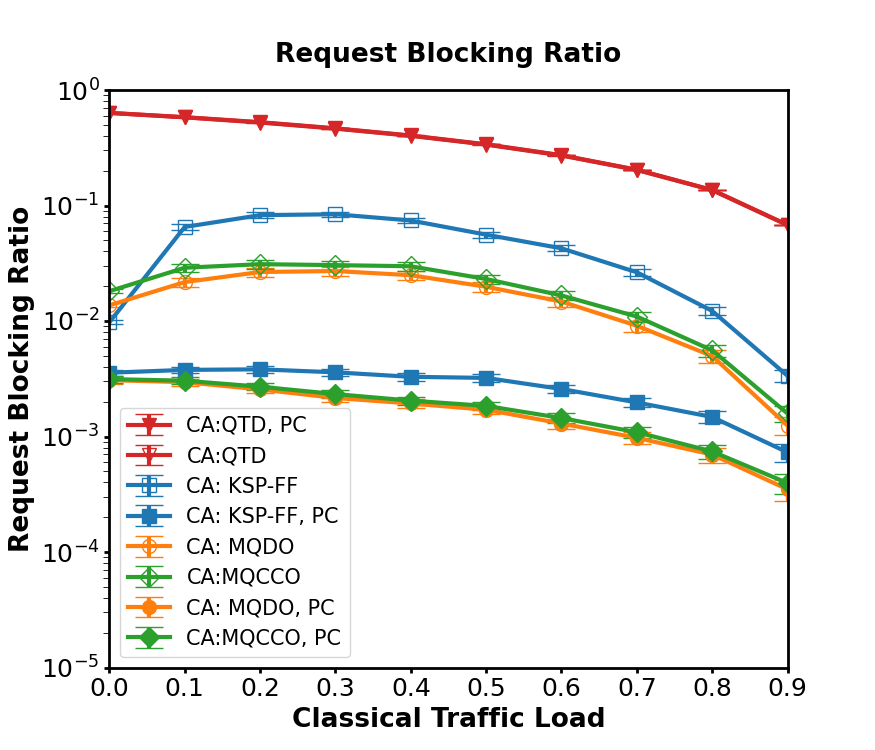}
		\caption{Blocking ratio in hybrid networks for $20$-node random Waxman topologies.}
		\label{fig:br_classical_waxman}
	\end{figure}

Finally, let us analyze the performance of Waxman topologies in hybrid quantum-classical networks. In this case, the blocking ratio is higher than that observed in Gabriel topologies, likely due to the structural properties of the Waxman graphs, which make it challenging to accommodate both quantum and classical channels while meeting the QSNR requirements. Interestingly, enabling the power control feature yields an improvement of up to one order of magnitude in blocking ratio—greater than the improvement observed with Gabriel topologies, as shown in Figure \ref{fig:br_classical_waxman}. Additionally, MQCCO and MQDO exhibit slightly better performance compared to the other algorithms in this scenario, which is a behaviour that was not observable in Gabriel topologies. These results suggest that certain network topologies may benefit from RWA algorithms tailored to their specific structural characteristics and resource constraints. Furthermore, increasing classical traffic initially raises the blocking ratio; however, once the load is high enough, the number of required quantum lightpaths decreases, facilitating more efficient resource utilization while meeting the QSNR requirements.

Overall, these findings highlight the influence of network topology on quantum-classical coexistence and demonstrate the importance of adaptive routing and power control strategies in optimizing performance.

\begin{figure}
		\centering
		\includegraphics[width=0.45\linewidth]{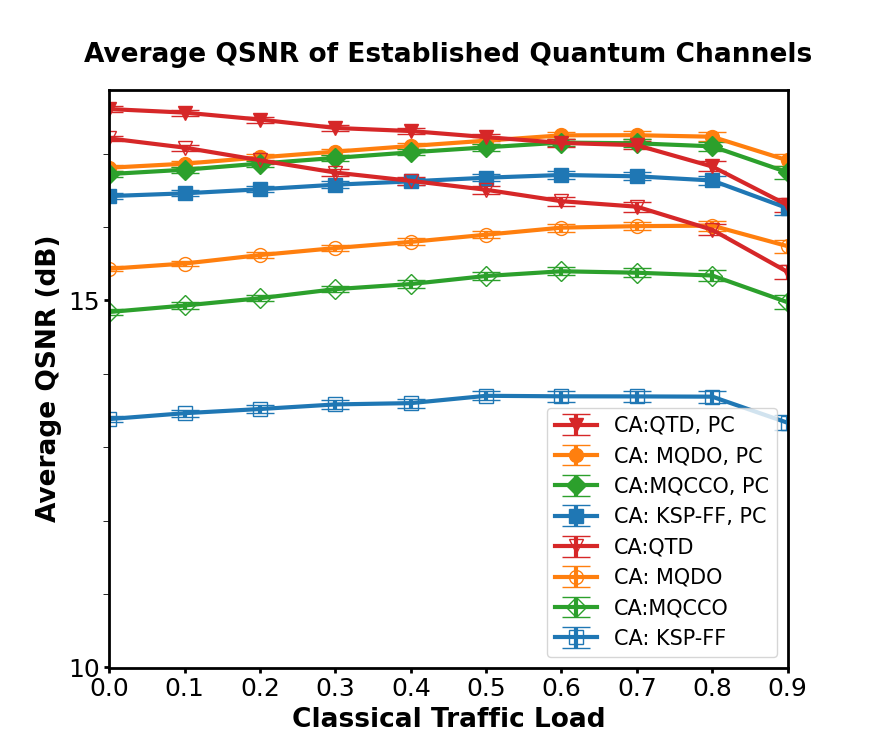}
		\includegraphics[width=0.45\linewidth]{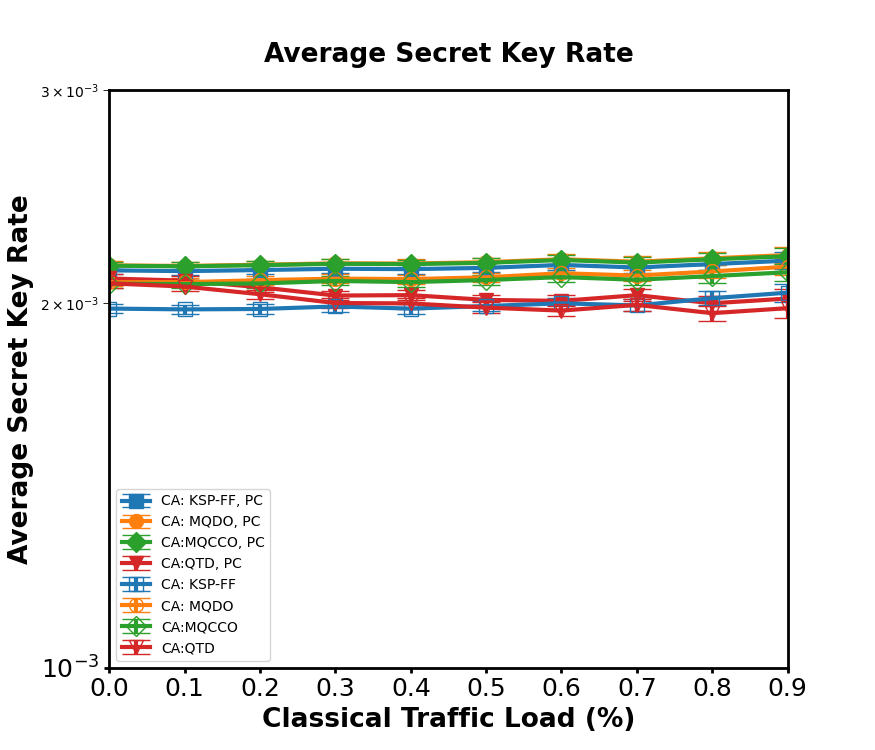}
		\caption{QSNR and SKR hybrid networks for $20$-node random Waxman topologies.}
		\label{fig:snr_kr_classical_waxman}
	\end{figure}

In terms of QSNR, Figure~\ref{fig:snr_kr_classical_waxman} (right) shows that SP-FF establishes QKD connections with the lowest signal quality. In contrast, MQDO, QTD, and MQCCO achieve QSNR values up to \SI{2}{\deci\bel} higher when power control is disabled. Enabling power control further enhances the quality of QKD signals by up to \SI{4}{\deci\bel} compared to the inactive case. In this scenario, the differences in QSNR among the algorithms become less significant when power control is active. Regarding the key generation rate, enabling power control results in a slight improvement in the generation speed.

Overall, these results confirm that, while power control has a positive effect on both QSNR and key generation rate, its impact tends to reduce algorithmic differences, making it a key enabler for more consistent QKD performance across routing strategies.

	\section{Conclusions}
	\label{Conclusions}
	In this study, we evaluate the performance of a power control algorithm over a traditional RWA algorithm like KSP-FF and other quantum-oriented RWA algorithms proposed in the literature. This protocol aims at reducing the launch power of the classical lightpaths in order to reduce the noise introduced to the quantum connections. Our study accounts for the undesired noise that classical channels potentially introduce into the quantum channels, proposing a QSNR formula inspired by the signal-to-interference-plus-noise ratio, SiNR, used in wireless networks. The study demonstrates that implementing a power control protocol effectively mitigates the noise introduced to quantum connections. This approach enhances the QSNR by reducing interference, thereby improving the quality of quantum channels.
	
	We have found that, in QKD networks, the application of the power control protocol alone can significantly improve the blocking ratio and enhances the quality of the established channels. This finding underscores the protocols's efficiency in quantum communication networks, where minimizing the interference of the classical channels is critical for maintaining high-quality secure communication.
	
	In hybrid traffic scenarios—where both quantum and classical connection requests coexist—we observed that network topology plays a critical role in performance. The studied graph models exhibited different sensitivities to interference and path availability, suggesting that topology-aware RWA strategies are essential. Moreover, as classical traffic increases, the cumulative effect of classical lightpaths can still degrade quantum channel quality, even with power control in place. In these cases, quantum-aware RWA algorithms such as MQDO, QTD, and MQCCO outperform general-purpose approaches by better accommodating the specific requirements of QKD traffic. Quantum-oriented RWA algorithms, which consider the unique requirements of quantum communication, are necessary to effectively manage network resources and maintain a high key rate. The proposed heuristics for routing and channel allocation minimize the impact of classical channels on quantum communication, ensuring the reliability and security of the network. However, to fully exploit their benefits, they should be assessed for dynamic traffic assignment and in combination with topology-aware and quantum-specific RWA strategies, which are left for future studies.
	
	Overall, our findings confirm that power control is a vital mechanism for enhancing QKD performance in mixed quantum-classical networks. The proposed routing and allocation heuristics contribute to minimizing the impact of classical interference, ensuring reliable and secure quantum communication even in complex, high-load network environments.
	
	\section*{Acknowledgments}
	Lidia Ruiz has been funded by the European Union NextGeneration UE/MICIU/Plan de  Recuperaci\'on, Transformaci\'on y Resiliencia/Junta de Castilla y Le\'on n and Junta de Castilla y Le\'on (Consejer\'ia de Educaci\'on/FEDER) Project VA184P24. J.C. Garcia-Escartin has been funded by the European Union NextGeneration UE/MICIU/Plan de Recuperaci\'on, Transformaci\'on y Resiliencia/Junta de Castilla y Le\'on and Junta de Castilla y Le\'on (Consejer\'ia de Educaci\'on/FEDER) Project VA184P24. Financial support of the Department of Education, Junta de Castilla y León, and FEDER Funds is gratefully acknowledged (Reference: CLU-2023-1-05).

\end{document}